\documentclass[12pt,a4paper]{article}
\usepackage[dvips]{pstcol}
\usepackage{amsbsy,latexsym,cite}
\usepackage{feynmp}
\usepackage{graphics}

\newcommand{\no}{\noindent}

\newcommand{\vb}{{\underline v}}

\fmfset{arrow_len}{5pt}\fmfset{arrpow_ang}{30}

\fmfcmd{%
  vardef middir(expr p,ang) =
   dir(angle direction length(p)/2 of p+ang)
  enddef;
  style_def arrow_left expr p =
    shrink(.7);
    cfill(arrow p shifted(4thick*middir(p,90)));
    endshrink
  enddef;
  style_def arrow_right expr p =
    shrink(.7);
    cfill(arrow p shifted(4thick*middir(p,-90)));
    endshrink
  enddef;}

\begin{document}
\setlength{\unitlength}{1mm}


\begin{center}
{\Large{\textbf{Charged Matter: Physics and
Renormalisation}\footnote{Talk presented by M.~Lavelle}}}
\end{center}
\bigskip
\begin{center}
\textsc{Emili Bagan\footnote{email: bagan@ifae.es}$^a$, Martin
Lavelle\footnote{email: m.lavelle@plymouth.ac.uk}$^b$ and David
McMullan}\footnote{email: d.mcmullan@plymouth.ac.uk}$^b$\\
[5truemm] $^a$\textit{Grup de F\'\i sica Te\`orica and IFAE\\
Universitat Aut\`onoma de Barcelona\\ E-08193 Bellaterra
(Barcelona)\\ Spain}
\\ [5truemm] $^b$\textit{School of Mathematics and Statistics\\
The University of Plymouth\\ Plymouth, PL4 8AA\\ UK}
\end{center}

\bigskip\bigskip\bigskip
\begin{quote}
\textbf{Abstract:} Descriptions of the ground state in unbroken
gauge theories with charged particles are discussed. In particular
it is shown that the on-shell Green's functions and $S$-matrix
elements corresponding to the scattering of these variables in QED
are free of soft and phase infra red divergences and that these
variables may be multiplicatively renormalised.
\end{quote}

\bigskip
\no \textbf{Introduction:}

\medskip

\noindent Everyone at this workshop is interested in constructing
gauge invariant variables, but not everything that is gauge
invariant is physically significant. Physics is generally
concerned with the lowest lying states, and in particular the
ground state. A gauge invariant \lq physical variable\rq\ may,
though, correspond to a highly excited state or even an infinitely
excited state and an infinitely excited state is not exactly
physical! Generally we do not know the form of the correct states.
In this talk we will investigate the behaviour of candidate lowest
energy states. In particular we will study these in the presence
of charges.

There are many physical reasons why we would like to be able to
have physical variables corresponding to charged particles. In the
non-abelian theory these are needed to understand how we may trace
the path from partons to constituent quarks. Their construction
could also be used to better understand how the creation of
(colour) charged particles generates colourless jets of hadrons.

A general formalism for constructing charged fields has already
been presented at this workshop (see D.~McMullan's talk). This
formalism relied on two inputs: local gauge invariance and a
further requirement called the dressing equation, which is
characterised by the well-defined velocity of the charged particle
a long time before or after a scattering process.

Although there are many powerful arguments in favour of these
candidate variables~\cite{Bagan:1999jf,Lavelle:1997ty}, these and
all other such descriptions {\it must} be checked in practical
calculations~\cite{Bagan:1997kg,Bagan:1999jk,d'Emilio:1999iu}. In
this talk we will specialise to the abelian theory and submit this
description to a variety of {\it perturbative} tests. We will, in
particular, examine the behaviour of the variables in both the
infra-red (IR) and the ultra-violet (UV) domains.

The IR problem is caused by erroneously identifying the Lagrangian
fermion with a physical charged particle. In the usual description
of any scattering process the on-shell Green's functions and the
$S$-matrix elements are plagued with infra-red divergences. (For
completeness we give a brief introduction to the IR problem
below.) We will see here that the appropriate use of the variables
we propose removes these divergences to all orders of perturbation
theory already at the level of matrix elements.

The need for controllable UV behaviour is immediately obvious and
we will demonstrate below both that the variables we use can be
multiplicatively renormalised and that standard physical results
(size of the electric charge, anomalous magnetic moment) can be
regained.

\bigskip

\no {\bf Charged Particles -- the Right Physical Variables}

\medskip

To construct charged particles we commence with the simple
observation that their physical description must be locally gauge
invariant. This requirement, however, is not strong enough to
single out a specific description. The next step is to note that
experimentally particles are characterised by their velocity. (We
stress that this only makes sense at asymptotic times; it would be
completely wrong to try to trace the path of a particle through
any scattering process.) This yields a further constraint, the
dressing equation, and with this we can construct the following
variables
\begin{equation}\label{vars}
h^{-1}\psi={\rm e}^{-ie K[A]}\;{\rm e}^{-ie \chi[A]}\psi \,,
\end{equation}
where
\begin{eqnarray}
         \chi [A] &
        = & {{\cal G}\cdot A\over{\cal G}\cdot\partial}  
        ;\qquad
        {\cal G} = (\eta+v)^\mu(\eta-v)\cdot\partial-\partial^\mu
        \\[0.5cm]
        K[A]&=&-\int_{\Gamma}(\eta+v)^\mu
        {\partial^\nu F_{\nu\mu}\over{\cal G}
        \cdot\partial}(x(s))ds \,;\quad \quad \eta^\mu  =
        (1,{\vec 0}) ;\;
        v^\mu  =  (0,\vb)\,,
\end{eqnarray}
We say that the matter field, $\psi$, has been {\it dressed} by
its appropriate electromagnetic cloud. This of course depends upon
its velocity and such a cloud is necessarily non-local.  The path
$\Gamma$ here depends on whether we are considering a charge
before or after scattering. It corresponds to extending backwards
to infinity the past trajectory (this is a straight line) of an
incoming charge. For outgoing charges we extend the straight line
corresponding to the future trajectory. This is quite different to
the path dependence of Mandelstam variables which is generally not
physically motivated, see~\cite{Lavelle:1999ki} on this.

We note that $\chi$ is gauge dependent and required for the
minimal condition of gauge invariance. It is a generalisation of
the variables introduced by Dirac~\cite{Dirac:1955ca} and
discussed by several speakers at this meeting. $K$ is gauge
invariant and is required, together with $\chi$, to fulfill the
dressing equation. We will see below that these two terms play
very different roles in the infra-red.

It might be objected that any particular $\chi$ can be set to zero
by an appropriate gauge choice. However, if we are considering
scattering where different particles have different momenta, then
different dressings must be associated to them -- no gauge choice
will then remove them all.

Two questions are immediately obvious: can we actually do anything
with these variables and are they useful? The rest of this talk is
devoted to showing that the answer to both is affirmative. This is
despite the necessarily non-local and non-covariant nature of the
variables.

\bigskip

\noindent \textbf{The Infra-Red Problem in Pair Creation}
\medskip

\no Before applying these variables, let us recall one of the
bugbears of unbroken gauge theories, the IR divergences which
occur in $S$-matrix elements and on-shell Green's functions. The
abelian problem may be seen in its most formidable fashion in the
context of the pair creation process. To keep the formulae simple,
we will consider scalar electrodynamics, the extension to the
fermionic theory does not bring any further insight since, for QED
with massive electrons, the IR problem is spin independent.

\bigskip
\begin{fmffile}{feyn}

The Feynman rules we require are (all loops involving the four
point vertex of scalar electrodynamics are IR finite):
\begin{eqnarray*}
 \parbox{20mm}{\begin{fmfgraph*}(20,15)
   \fmfleft{i}\fmfright{o}\fmf{fermion,label=$^p$}{i,o}
 \end{fmfgraph*}}
 &=&\frac{i}{p^2-m^2+i\varepsilon}\\
\parbox{24mm}{\begin{fmfgraph*}(24,15)
   \fmfleft{i}\fmfright{o}\fmf{photon,label=$^k$}{i,o}
   \fmflabel{$\mu$}{i}
   \fmflabel{$\nu$}{o}
 \end{fmfgraph*}}\
 &=&-i\frac{g^{\mu\nu}}{k^2+i\varepsilon}\\
 \parbox{20mm}{\begin{fmfgraph*}(20,15)
   \fmfleft{i}\fmfright{o}\fmftop{g}\fmf{fermion,label=$^p$}{i,v}
   \fmf{fermion,label=$^{p'}$,label.side=right}{v,o}
   \fmf{photon,tension=0,label=$^\mu$}{g,v}
 \end{fmfgraph*}}
 &=&ie(p+p')_\mu\,.
\end{eqnarray*}

\no Although we here work in Feynman gauge, all results for our
gauge invariant variables will be visibly gauge invariant as are
the usual $S$-matrix elements\footnote{The on-shell Green's
functions of the Lagrangian matter fields in general display IR
divergences. The IR-finiteness of the propagator in Yennie gauge
does {\it not} hold for other Green's functions.}.

Applying these to the pair creation process (we may set the vertex
factor to be unity for simplicity)
\medskip
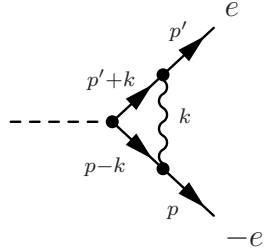
\begin{center}
    \begin{fmfgraph*}(30,25)
      \fmfleft{source}
      \fmf{dashes}{source,ver}
      \fmf{fermion,label=$^{{p-k}}$}{ver,v1}
      \fmf{fermion,label=$^p$}{v1,pone}
      \fmf{fermion,label=$^{p'+k}$,label.dist=.5pt}{ver,v2}
\fmf{fermion,label=$^{p'}$,label.side=left,label.dist=.5pt}{v2,ptwo}
      \fmf{photon,tension=0,label=$^k$}{v1,v2}
      \fmfright{pone,ptwo}
      \fmfdot{ver,v1,v2}
      \fmflabel{$-e$}{pone}
      \fmflabel{$e$}{ptwo}
    \end{fmfgraph*}
\end{center}
This yields

\begin{eqnarray*}
&&\frac{-ie^2}{(p^2-m^2)({p'}^2-m^2)}
\int\!{d}^4k\,
\frac{g_{\mu\nu}}{(k^2+i\varepsilon)}\times\\&&\,\,\,\,\,
\frac{(2p'+k)^\mu(2p-k)^\nu}{
\left[(p'+k)^2-m^2+i\varepsilon\right]
\left[(p-k)^2-m^2+i\varepsilon\right]}\,.
\end{eqnarray*}

 To calculate the $S$-matrix, we need to go on-shell and
calculate the residues of the two poles (one per external leg).
The IR problem shows itself in such residues, both here and also
in the Green's functions themselves should we choose the on shell
renormalisation scheme. Dropping all IR finite terms we obtain for
the residue
\begin{equation}
\int\! {d}^4k\, \frac{g_{\mu\nu}{p'}^\mu p^\nu}{(p'\cdot
k+i\varepsilon) (p\cdot k-i\varepsilon)(k^2+i\varepsilon)}\,.
\end{equation}
We may integrate over $k_0$ and pick up the poles.
Note that there are two different types of divergences here.
\begin{itemize}
  \item From $k^2+i\varepsilon=0$ we get the
  so-called {\it soft divergences}, which
here contribute
  ($v$ being the relative velocity)
  \[
\frac1{8\pi^2}\ln\left(\frac{\Lambda}{\lambda}\right)\frac1{|v|}
\ln\left(\frac{1+|v|}{1-|v|}\right)
 \]

  \item From the poles at $p\cdot k-i\varepsilon=0$ and $p'\cdot
k+i\varepsilon=0$, we obtain the so-called {\it phase divergences}
which here contribute
\[
-i\frac1{4\pi}\frac1{|v|}\ln\left(\frac{\Lambda}{\lambda}\right)
 \]

\end{itemize}

Such phase divergences are imaginary and so only enter in the
unobservable phase and are therefore ignored in many treatments of
QED. Similar structures are though important in QCD. Note that
were we to have considered a scattering process there would not be
any phase divergence as these poles would all be in one half plane
and could be avoided.

Other contributions to the soft divergence in the pair creation
process come from

\bigskip
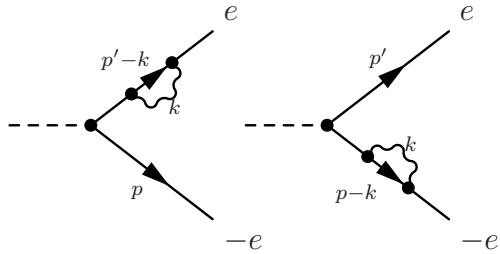
\begin{center}
    \begin{fmfgraph*}(30,25)
      \fmfleft{source}
      \fmf{dashes}{source,ver}
      \fmf{fermion,label=$_{{p}}$,tension=1/3}{ver,pone}
      \fmf{plain}{ver,v1}
      \fmf{plain}{v2,ptwo}
      \fmf{fermion,label=$^{p'-k}$,label.dist=.5pt}{v1,v2}
      \fmf{photon,right,tension=0,label=$_k$,label.dist=.5pt}{v1,v2}
      \fmfright{pone,ptwo}
      \fmfdot{ver,v1,v2}
      \fmflabel{$-e$}{pone}
      \fmflabel{$e$}{ptwo}
    \end{fmfgraph*}
    \begin{fmfgraph*}(30,25)
       \fmfleft{source}
      \fmf{dashes}{source,ver}

\fmf{fermion,label=$^{{p'}}$,label.dist=.5pt,tension=1/3}{ver,ptwo}
      \fmf{plain}{ver,v1}
      \fmf{plain}{v2,pone}
      \fmf{fermion,label=$_{p-k}$}{v1,v2}
      \fmf{photon,left,tension=0,label=$_k$,label.dist=.5pt}{v1,v2}
      \fmfright{pone,ptwo}
      \fmfdot{ver,v1,v2}
      \fmflabel{$-e$}{pone}
      \fmflabel{$e$}{ptwo}
    \end{fmfgraph*}
\end{center}
\bigskip

These IR divergences mean that in unbroken gauge theories we
seemingly  cannot talk about matrix elements or on-shell Green's
functions! However, we have been using unphysical variables and so
maybe this is not so surprising.

\bigskip

\no \textbf{Dressings and the Infra-Red}

\medskip

It is generally understood that the IR problem results from the
neglect of asymptotic interactions (we cannot just switch off the
coupling, see R.~Horan's talk at this workshop
and~\cite{Horan:1999ba}). We have argued
previously~\cite{Bagan:1999jf} that our dressed fields have the
appropriate interactions incorporated into them, so that the
coupling \textit{does} effectively switch off and so we expect to
be able to carry out the LSZ programme without encountering IR
singularities.

To calculate dressed Green's functions we expand the dressings
themselves in the coupling constant. The two terms in the dressing
so generate the new rules:
\bigskip

\begin{eqnarray*}
{
\chi:}\quad\fmfcmd{%
  vardef middir(expr p,ang) =
   dir(angle direction length(p)/2 of p+ang)
  enddef;
  style_def arrow_left expr p =
    shrink(.7);
    cfill(arrow p shifted(4thick*middir(p,90)));
    endshrink
  enddef;
  style_def arrow_right expr p =
    shrink(.7);
    cfill(arrow p shifted(4thick*middir(p,-90)));
    endshrink
  enddef;}
 \parbox{20mm}{\begin{fmfgraph*}(20,20)
   \fmfleft{i}\fmfright{o}\fmftop{g}\fmf{fermion}{i,o}
   \fmf{photon,tension=0,label=$k$,label.side=left}{g,o}
   \fmflabel{$\mu$}{g}
   \fmf{arrow_right}{g,o}
   \fmfv{decor.shape=circle,decor.filled=empty,decor.size=3pt}{o}
 \end{fmfgraph*}}
 &=&\frac{eV^\mu}{V\cdot k}\\
 {
 K:}\quad\parbox{20mm}{\begin{fmfgraph*}(20,20)
   \fmfleft{i}\fmfright{o}\fmftop{g}\fmf{fermion,label=$p- k$}{i,o}
   \fmf{photon,tension=0,label=$k$,label.side=left}{g,o}
   \fmflabel{$\mu$}{g}
   \fmf{arrow_right}{g,o}
   \fmfv{decor.shape=cross,decor.size=5pt}{o}
 \end{fmfgraph*}}
 &=&\frac{eW^\mu}{V\cdot k}
\end{eqnarray*}

\medskip

\no where $V^\mu=(\eta+v)^\mu(\eta-v)\cdot k-k^\mu$ and
\begin{equation}
W^\mu=\frac{k^\mu(\eta+v)\cdot k-(\eta+v)^\mu
k^2}{k\cdot(\eta+v)-i\varepsilon}\,.
\end{equation}
Note that $k\cdot W=0$  is an expression of the gauge invariance
of $K$. We stress that $V\cdot k$ is not singular since
$|\underline{v}|<1$.

Now let us return to the pair production process. If physical
(dressed) fields are being produced we have extra Feynman
diagrams. We find that soft divergences also arise in the
following diagram

\bigskip

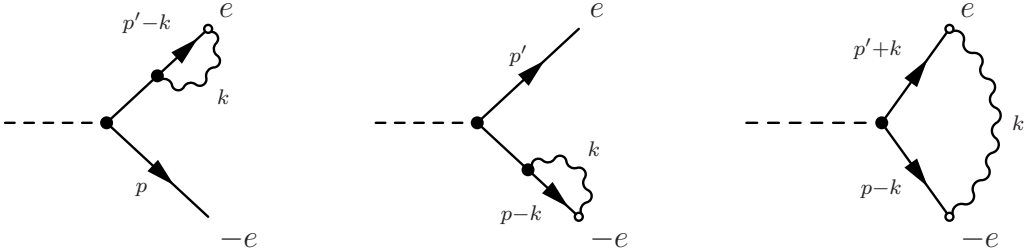
\begin{center}
    \begin{fmfgraph*}(30,25)
      \fmfleft{source}
      \fmf{dashes}{source,ver}
      \fmf{fermion,label=$_{{p}}$,tension=1/2}{ver,pone}
      \fmf{plain}{ver,v1}
      \fmf{fermion,label=$^{p'-k}$,label.side=left}{v1,ptwo}
      \fmf{photon,right,tension=0,label=$_k$}{v1,ptwo}
      \fmfright{pone,ptwo}
      \fmfdot{ver,v1}

\fmfv{decor.shape=circle,decor.filled=empty,decor.size=3pt}{ptwo}
      \fmflabel{$-e$}{pone}
      \fmflabel{$e$}{ptwo}
    \end{fmfgraph*}
    $\qquad\qquad$
    \begin{fmfgraph*}(30,25)
       \fmfleft{source}
      \fmf{dashes}{source,ver}

\fmf{fermion,label=$^{{p'}}$,label.dist=.5pt,tension=1/2}{ver,ptwo}
      \fmf{plain}{ver,v1}
      \fmf{fermion,label=$_{p-k}$,label.side=right}{v1,pone}
      \fmf{photon,left,tension=0,label=$_k$}{v1,pone}
      \fmfright{pone,ptwo}
      \fmfdot{ver,v1}

\fmfv{decor.shape=circle,decor.filled=empty,decor.size=3pt}{pone}
      \fmflabel{$-e$}{pone}
      \fmflabel{$e$}{ptwo}
  \end{fmfgraph*}
$\qquad\qquad$
   \begin{fmfgraph*}(30,25)
       \fmfleft{source}
      \fmf{dashes}{source,ver}
      \fmf{fermion,label=$^{{p'+k}}$,label.side=left}{ver,ptwo}
      \fmf{fermion,label=$_{p-k}$,label.side=right}{ver,pone}
      \fmf{photon,right=.5,tension=0,label=$_k$}{pone,ptwo}
      \fmfright{pone,ptwo}
      \fmfdot{ver}
\fmfv{decor.shape=circle,decor.filled=empty,decor.size=3pt}{pone,ptwo}
      \fmflabel{$-e$}{pone}
      \fmflabel{$e$}{ptwo}
    \end{fmfgraph*}

\end{center}

\bigskip
\noindent Other diagrams do not yield soft
divergences\footnote{This is a slight oversimplification: some of
the diagrams involving the additional $K$~part of the dressing
also generate soft divergences, the apparent divergences of this
subset of graphs sum to zero among themselves.}.

Calculating these diagrams along the lines discussed above, we
obtain the following total soft contribution to the residue:
\begin{eqnarray*}
\int\! d^4k\,&&\!\!\!\!\!\!\!\! \left\{ \left(\frac{p^\mu}{p\cdot
k}-\frac{{V'}^\mu}{{V'}\cdot k}\right) \frac{g_{\mu\nu}}{k^2}
\left(\frac{V^\nu}{V\cdot k}-\frac{{p'}^\nu}{{p'}\cdot
k}\right)\right.\\ && \qquad -\left.\left(\frac{p^\mu}{p\cdot
k}-\frac{{p'}^\mu}{{p'}\cdot k}\right) \frac{g_{\mu\nu}}{k^2}
\left(\frac{p^\nu}{p\cdot k}-\frac{{p'}^\nu}{{p'}\cdot k}\right)
\right\}
\end{eqnarray*}

\bigskip

From this result we see that gauge invariance is manifest.
(Replacing the Feynman gauge propagator ($\sim g^{\mu\nu}$) by any
more general one will always introduce a $k^\mu$ or a $k^\nu$
which will vanish when dotted into the round brackets above.

Although we can show the IR-finiteness of this result by brute
force, this cancellation of soft divergences can be essentially
read off upon realising that in this integral we can replace
$V^\mu\to u^\mu$, ${V'}^\mu\to {u'}^\mu$ in the soft region where
$k^2\approx0$. Thus these divergences completely
  vanish when we go on-shell and evaluate the residues at the
  appropriate points on the mass shell,
  $p^\mu=m u^\mu$ and ${p'}^\mu=m{u'}^\mu$, which were used as
  fundamental inputs in our
  construction of the dressings, $\chi$.
  Since as is well known IR divergences in QED exponentiate, it is
  not completely surprising that this argument can in fact be
  extended~\cite{Bagan:1997kg,Bagan:1999jk} and we have shown
  that the soft divergences cancel to all orders in
perturbation theory.

We note that this is a very subtle test: if the velocity parameter
in the dressing is not matched to the renormalisation point the IR
divergences will not cancel! This is because the dressing  then no
longer describes the electromagnetic cloud corresponding to the
lowest lying state of a charged particle with that velocity.

In a similar manner, the phase divergence seen earlier is
cancelled by the contribution from the other ($K$) part of the
dressing. This is generated by one specific diagram, viz
\medskip
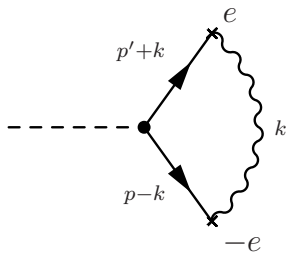
\begin{center}
    \begin{fmfgraph*}(30,25)
       \fmfleft{source}
      \fmf{dashes}{source,ver}
      \fmf{fermion,label=$^{{p'+k}}$,label.side=left}{ver,ptwo}
      \fmf{fermion,label=$_{p-k}$,label.side=right}{ver,pone}
      \fmf{photon,right=.5,tension=0,label=$_k$}{pone,ptwo}
      \fmfright{pone,ptwo}
      \fmfdot{ver}
      \fmfv{decor.shape=cross,decor.size=5pt}{pone,ptwo}
      \fmflabel{$-e$}{pone}
      \fmflabel{$e$}{ptwo}
    \end{fmfgraph*}
\end{center}

\no for more detail, see~\cite{Horan:1998im}.

We thus see that all of the IR divergences of QED are removed if
we use our dressed fields and renormalise them at the right point
on the mass shell. This gives us a great deal of confidence in the
physical interpretation of these variables.

\bigskip
\goodbreak

\no \textbf{UV Structure}

\medskip

\no We now want to study the UV behaviour of these variables.
Since they are non-covariant we might be worried, especially if we
think of the notorious problems with axial gauges, that there
could be problems such as non-multiplicative renormalisability.

In fact such fears are unfounded. We have seen that the UV
behaviour of these variables is excellent as we now describe. Note
that in what follows we have solely used the minimal dressing
$\chi$.

In the usual propagator there are two renormalisation constants:
the mass shift and the wave function renormalisation constant. The
former is gauge invariant (and IR finite), the latter is usually
IR divergent in an on-shell scheme. For the dressed fields the
mass shift is unchanged and we find that multiplicative wave
function renormalisation is possible.

In scalar QED at one loop, this renormalisation constant, $\phi\to
\sqrt{Z_2}\phi$, can be found~\cite{Bagan:1997dh} to be
\begin{eqnarray}
Z_2&\!\!=\!\!& 1+\frac\alpha{4\pi}\left\{
(6+2\chi(v))\frac1{{\hat{\epsilon}}} +4\left(
1-\gamma^{-2}\chi(v)-{\phantom{\frac11}}\right.\right.\nonumber\\
&& \left.\left.\quad\qquad \frac1{|{\vb}|}\left[
L_2(|{\vb}|)-L_2(-|{\vb}|) \right] \right)\right\}
\end{eqnarray}
where $L_2$ is the dilogarithm and $\chi({\vb})={\vert
{\vb}\vert}^{-1}{\rm ln}\left\{\left(1-\vert{\vb}
\vert\right)/(1+\vert{\vb}\vert)\right\} $. This renormalisation
constant is of course IR finite.

For fermionic QED: IR finite wave function renormalisation is also
possible. This though turns out to be a  matrix
multiplication~\cite{Bagan:1997su}. This is, we believe, linked
with the interplay between Lorentz boosts and gauge
transformations in the charged sector.  For details see Sect.~8
of~\cite{Lavelle:1997ty}. We have also studied the renormalisation
of $\psi_v$ when it is understood as a composite
  operator. This turns out~\cite{Bagan:1999jk}
  to be very  well behaved. Almost all
  diagrams are UV-finite and we have shown that these variables do
{\it  not} mix with each other. This multiplicative
renormalisation of the dressed matter fields and the non-mixing of
the operators (the renormalisation of composite operators is, we
recall, often plagued by such mixing) are very strong evidence
that we are correct in identifying our variables with physical
degrees of freedom.

We have further constructed a Ward identity for the vertex
describing the scattering of charged particles. This has been
shown to hold in explicit calculations.

A further renormalisation constant is found necessary when we
study  vertex renormalisation for non-trivial scattering. This
constant is just that which occurs in the Isgur-Wise function and
in the renormalisation of Wilson loop cusps. In other words this
is the renormalisation constant which occurs when a charge is
suddenly scattered. Its appearance in our formalism is therefore
highly welcome: it signals the renormalisation needed when a
charge is accelerated and the electromagnetic fields around the
charge (the dressing) need to be rearranged.

We have tested that some of the usual physical predictions of QED
hold. In particular we have seen that the standard one-loop
prediction for the anomalous magnetic moment $g-2$ again emerges
if dressed matter is used and that the value of the charge is
again obtained.

Finally, we conjecture that the good IR behaviour of our fields
may help us to extract some physical predictions from QED.
Quantities like the charge radius which are usually IR divergent
(since $F_1(q^2\neq 0)$ displays such divergences) may become
finite.

\bigskip

\no \textbf{Conclusions}

\medskip

\no There are various gauge invariant variables on the market.
Generally  though they do not have any clear physical meaning. The
Wilson loop route to the interquark potential makes a virtue of
this by solely requiring that the state which is evolved (two
heavy colour sources linked by a string) has a non-zero overlap
with the ground state. Here we have set ourselves a more ambitious
goal and argued that a specific set of variables, in the charged
sector, has a specific physical interpretation: namely they
correspond to charged particles with well defined velocities in
the asymptotic region before or after scattering.

We have recalled that the usual LSZ route to the $S$-matrix
displays for gauge theories IR divergences. We have seen that our
variables, which have a structured form, remove these divergences
already at the level of matrix elements or on-shell Green's
functions. This cancellation requires an exact correspondence
between the velocity parameter in the dressing and the point on
the mass shell where we renormalise it.

The UV behaviour of these fields has also been tested.
Multiplicative renormalisation; no operator mixing when we
consider the fields as composite operators; Ward identities and
the reproduction of standard physical results --- all these argue
for the variables we have introduced.

In the non-abelian theory we further recall that the extension of
the minimal dressing $\chi$ to order $g^4$ has been shown to yield
the anti-screening component of the interquark potential.

All of these results give us confidence that this programme has a
sound physical basis. What then are the next steps?

In QCD a new type of \lq infra-red\rq\ divergences, collinear
singularities, arise as a consequence of the masslessness of the
gluon. The simplest way to study these divergences is to let the
electron mass be zero. We need then to solve the dressing equation
in this subtle limit~\cite{Horan:1998im} and perturbatively
calculate the resulting Green's functions.

There are two challenges in massive QED. We still want to carry
out full studies of QED cross-sections using these variables and
study quantities such as the charge radius. Another question
though is can we consider bound states such as positronium? A
possible tool here would be Haag's
expansion~\cite{Greenberg:1994zu}, where we would urge the use of
the physical electron fields as the asymptotic states
corresponding to individual particles.

\end{fmffile}

\bigskip
\no\textbf{Acknowledgements:} This work was supported by the
British Council/Spanish Education Ministry \textit{Acciones
Integradas} grant no.\ Integradas grant 1801 /HB1997-0141. It is a
pleasure to thank the local organisers for their hospitality and
PPARC for travel support.

%
%

\begin{thebibliography}{10}

\bibitem{Bagan:1999jf}
E.~Bagan, M.~Lavelle, and D.~McMullan,
\newblock (1999), hep-ph/9909257,
\newblock \textit{Charges from
Dressed Matter: Construction}, BNL-HET-99/18,
  PLY-MS-99-23, submitted for publication.

\bibitem{Lavelle:1997ty}
M.~Lavelle and D.~McMullan,
\newblock Phys. Rept. {\bf 279}, 1 (1997), hep-ph/9509344.

\bibitem{Bagan:1997kg}
E.~Bagan, M.~Lavelle, and D.~McMullan,
\newblock Phys. Rev. {\bf D57}, 4521 (1998), hep-th/9712080.

\bibitem{Bagan:1999jk}
E.~Bagan, M.~Lavelle, and D.~McMullan,
\newblock (1999), hep-ph/9909262,
\newblock \textit{Charges from
Dressed Matter: Physics and Renormalisation},
  BNL-HET-99/19, PLY-MS-99-24, submitted for publication.

\bibitem{d'Emilio:1999iu}
E.~d'Emilio and S.~Micciche,
\newblock (1999), hep-th/9908119,
\newblock \textit{Infrared asymptotic
dynamics of gauge invariant charged
  fields: QED versus QCD}.

\bibitem{Lavelle:1999ki}
M.~Lavelle and D.~McMullan,
\newblock (1999), hep-ph/9910398,
\newblock \textit{Hadrons Without
Strings}, to appear in Physics Letters B.

\bibitem{Dirac:1955ca}
P.~A.~M. Dirac,
\newblock Can. J. Phys. {\bf 33}, 650 (1955).

\bibitem{Horan:1999ba}
R.~Horan, M.~Lavelle, and D.~McMullan,
\newblock (1999), hep-th/9909044,
\newblock \textit{Asymptotic
Dynamics in Quantum Field Theory}, submitted for
  publication.

\bibitem{Horan:1998im}
R.~Horan, M.~Lavelle, and D.~McMullan,
\newblock Pramana J. Phys. {\bf 51}, 317 (1998), hep-th/9810089,
\newblock Erratum-ibid, 51 (1998) 235.

\bibitem{Bagan:1997dh}
E.~Bagan, B.~Fiol, M.~Lavelle, and D.~McMullan,
\newblock Mod. Phys. Lett. {\bf A12}, 1815 (1997), hep-ph/9706515,
\newblock Erratum-ibid, A12 (1997) 2317.

\bibitem{Bagan:1997su}
E.~Bagan, M.~Lavelle, and D.~McMullan,
\newblock Phys. Rev. {\bf D56}, 3732 (1997), hep-th/9602083.

\bibitem{Greenberg:1994zu}
O.~W. Greenberg,
\newblock (1994), hep-ph/9502253,
\newblock \textit{Virtues of the
Haag expansion in quantum field theory}.

\end{thebibliography}

\end{document}